\documentclass[aps,showpacs,superscriptaddress,preprint]{revtex4}%
\usepackage{amsfonts}
\usepackage{amsmath}
\usepackage{amssymb}
\usepackage{multirow}
\usepackage{graphicx}%
\providecommand{\U}[1]{\protect\rule{.1in}{.1in}}
\providecommand{\U}[1]{\protect\rule{.1in}{.1in}}
\providecommand{\U}[1]{\protect\rule{.1in}{.1in}}
\providecommand{\U}[1]{\protect\rule{.1in}{.1in}}
\providecommand{\U}[1]{\protect\rule{.1in}{.1in}}
\providecommand{\U}[1]{\protect\rule{.1in}{.1in}}

\begin{document}
\title{
Influences of Al doping on the electronic structure of Mg(0001) and
dissociation property of H$_2$}

\author
{Yanfang Li}

\affiliation{College of Materials Science and Engineering, Taiyuan
University of Technology,Taiyuan 030024, People's Republic of China}

\affiliation{LCP, Institute of Applied Physics and Computational
Mathematics, P.O. Box 8009,Beijing 100088, People's Republic of
China}

\author
{Yu Yang}

\affiliation{LCP, Institute of Applied Physics and Computational
Mathematics, P.O. Box 8009,Beijing 100088, People's Republic of
China}

\author
{Yinghui Wei}

\affiliation{College of Materials Science and Engineering, Taiyuan
University of Technology,Taiyuan 030024, People's Republic of China}

\author
{Ping Zhang}
\thanks{Corresponding author.
E-mail address: zhang\underline{ }ping@iapcm.ac.cn}

\affiliation{LCP, Institute of Applied Physics and Computational
Mathematics, P.O. Box 8009,Beijing 100088, People's Republic of
China}

\date{\today}

\begin{abstract}

By using the density functional theory method, we systematically
study the influences of the doping of an Al atom on the electronic
structures of the Mg(0001) surface and dissociation behaviors of
H$_2$ molecules. We find that for the Al-doped surfaces, the surface
relaxation around the doping layer changes from expansion of a clean
Mg(0001) surface to contraction, due to the redistribution of
electrons. After doping, the work function is enlarged, and the
electronic states around the Fermi energy have a major distribution
around the doping layer. For the dissociation of H$_2$ molecules, we
find that the energy barrier is enlarged for the doped surfaces.
Especially, when the Al atom is doped at the first layer, the energy
barrier is enlarged by 0.30 eV. For different doping lengths,
however, the dissociation energy barrier decreases slowly to the
value on a clean Mg(0001) surface when the doping layer is far away
from the top surface. Our results well describe the electronic
changes after Al-doping for the Mg(0001) surface, and reveal some
possible mechanisms for improving the resistance to corrosion of the
Mg(0001) surface by doping of Al atoms.

\end{abstract}


\maketitle

\section{Introduction}

Since hydrogen is one of the best clean fuels in the future, looking
for an ideal storage material has become an important task for
experimental and theoretical researchers. Among all the candidates,
magnesium (Mg) based materials are widely concerned because they are
relatively inexpensive and can hold a high weight percentage of
hydrogen \cite{Fritzsche08,Grochala04}. So the interaction between
hydrogen and Mg has been extensively studied
\cite{Norskov1981,Sprunger1991,Bird1993,Vegge04,Johansson06,Wu08}.
The main disadvantage that prevents the vast applications of Mg for
hydrogen storage lies in that the corresponding hydrogenation and
dehydrogenation temperatures are large
\cite{Fritzsche08,Grochala04}. Recently, it is found that the
aluminum (Al) doped MgAl thin films have much lower hydrogenation
temperatures than pure Mg films \cite{Fritzsche08,Fritzsche09}.
However, the specific reason for the improvement as well as the
influence of Al doping on the electronic properties of Mg films,
remains unclear. So, in order to advance the searches for ideal
hydrogen storage materials and applications of hydrogen fuels, it
become quite important to make clear the influence of Al doping on
the electronic structure of Mg and the the interaction between
hydrogen and Mg.

Apart from the potential usages in hydrogen storage, MgAl alloys are
also important industrial materials used in aerospace applications
\cite{Kainer2000}. It is always believed that the Al doping makes Mg
more resistent to being corrupted by air. So lots of efforts have
been applied to upgrade the manufacturing techniques to dope Al in
Mg materials, and now the technology has become quite mature.
Studying the interactions between small molecules and Al-doped Mg
surfaces can give us critical information about the mechanisms for
Al doping to improve the resistance of Mg to corruption, and thus
are very meaningful and necessary. Based on this background and the
requirements of hydrogen storage materials, we here perform first
principles calculations to systematically study the influence of Al
doping on the electronic properties of the Mg(0001) surface and the
adsorption behavior of hydrogen molecules.

Previous studies have revealed that the Mg(0001) surface has both
considerable $s$ and $p$ electronic states distributing around the
Fermi energy, because of the $sp$ hybridizations \cite{Li09}. Here
we further find that after doping of an Al atom, the electronic
structures around the Fermi energy mainly distribute around the
doping layer. For the dissociation of hydrogen molecules on the
clean Mg(0001) surface, it has been found that the most
energetically favored site is the surface bridge site
\cite{Vegge04,Johansson06,Wu08,Li09CPB}, with the corresponding
minimum energy barrier of 0.85 eV \cite{Li09CPB}. Through our
present first-principles calculations, we find that Al-doping will
not change the most energetically favored dissociation channel.
However, the minimum energy barrier for dissociation of hydrogen is
critically dependent on the doping depth of Al, and the values are
all larger than that on the clean Mg(0001) surface. The rest of this
paper is organized as follows: In Sec. II, we describe our
fist-principles calculation method and the models used in this
paper. In Sec. III, we present in detail our calculated results,
including the comparisons of the electronic structures of the
Mg(0001) surface before and after Al doping, and the dissociation of
hydrogen molecules on the Al-doped Mg(0001) surface. At last, the
conclusion is given in Sec. IV.

\section{Calculational methods}

Our calculations are performed within density functional theory
(DFT) using the Vienna {\it ab-initio} simulation package (VASP)
\cite{VASP}. The PW91 \cite{PW91} generalized gradient approximation
and the projector-augmented wave potential \cite{PAW} are employed
to describe the exchange-correlation energy and the electron-ion
interaction, respectively. The cutoff energy for the plane wave
expansion is set to 250 eV, which is large enough to make the
calculational error of the adsorption energy below 0.01 eV. The
clean and Al-doped Mg(0001) surface are modeled by a slab composing
of five atomic layers and a vacuum region of 20 \AA. The $2\times2$
supercell in which each monolayer contains four Mg atoms are adopted
in the study of the H$_2$ adsorption. Our test calculations have
shown that the $2\times2$ supercell is sufficiently large to avoid
the interaction between adjacent hydrogen molecules. Integration
over the Brillouin zone is done using the Monkhorst-Pack scheme
\cite{Monkhorst} with $11\times11\times1$ grid points. A Fermi
broadening \cite{Weinert1992} of 0.1 eV is chosen to smear the
occupation of the bands around the Fermi energy (E$_f$) by a
finite-$T$ Fermi function and extrapolating to $T=0$ K. During
geometry optimizations, the bottom layer of the clean and Al-doped
Mg(0001) surface is fixed while other Mg and Al atoms are free to
relax until the forces on them are less than 0.01 eV/\AA. The
calculation of the potential energy surface for molecular H$_2$ is
interpolated to 209 points with different bond length ($d_{\rm
H-H}$) and height ($h_{\rm H_2}$) of H$_{2}$ at each surface site.
The calculated lattice constant of bulk Mg ($a$, $c$) and the bond
length of a free H$_{2}$ molecule are 3.21 \AA, 5.15 \AA~ and 0.75
\AA, respectively, in good agreement with the experimental values of
3.21 \AA, 5.20 \AA~\cite{Amonenko1962,Ashcroft1976} and 0.74 \AA~\
\cite{Huber1979}.

\section{Results and discussion}

As shown in Fig. 1, we have employed 3 different doping depths for
Al in the Mg(0001) surface, with the Al atom respectively in the
first (MgAl1), second (MgAl2) and third monolayer (MgAl3). For the
clean Mg(0001) surface, our calculations reveal that the distances
between the three topmost atomic layers are expanded from their bulk
values, in accordance with previous theoretical
\cite{Staikov1999,Wachowicz01,Wu08} and experimental reports
\cite{Rotenberg2000}. It is because that due to its special surface
charge redistribution, the topmost three layers of the Mg(0001)
surface are negatively charged and hence repel each other. In fact,
the surface expansion is a special phenomena for the Mg(0001)
surface, which is not seen for other surfaces of Mg
\cite{Staikov1999}. And instead of expansions, the surface
relaxations of many metals show contractions \cite{Wu08}. The
relaxation calculation for the clean Mg(0001) surface also confirms
the precise of our methods.

\begin{figure}[ptb]
\includegraphics[angle=0,width=1.0\textwidth]{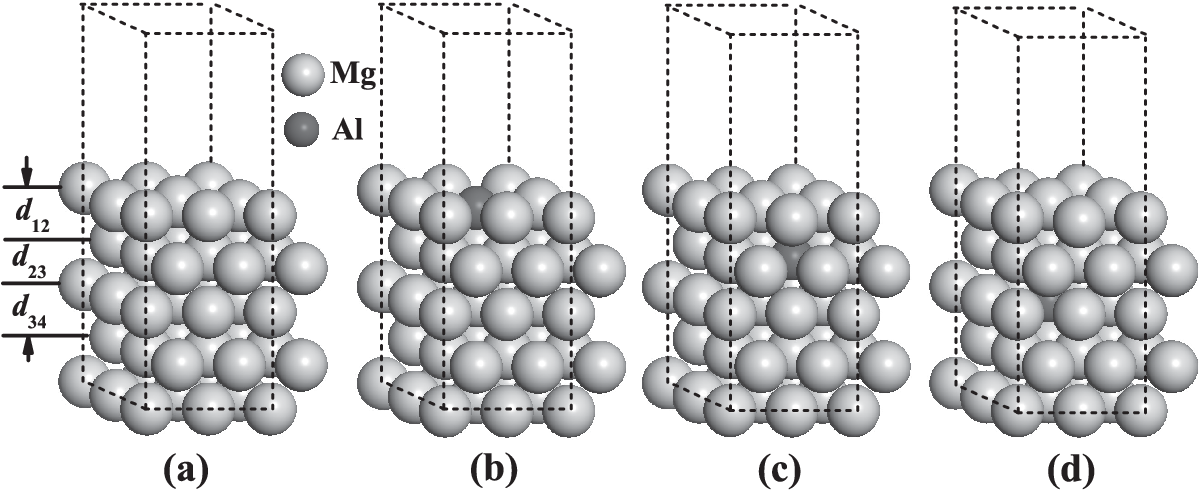}
\caption{(Color online). Atomic configurations of the Mg(0001) (a),
MgAl1 (b), MgAl2 (c), and MgAl3 (d) surfaces, with the corresponding
supercells depicted in dashed lines. Mg and Al atoms are shown in
grey and dark grey balls.}
\label{fig1}%
\end{figure}

However, after the doping of an Al atom, the surface relaxation is
changed. To describe more clearly the changes, here we define the
surface relaxation as
\begin{equation}\label{Dd}
\Delta d_{ij}=(d_{ij}-d_0)/d_0,
\end{equation}
where $d_{ij}$ and $d_0$ are respectively the distance between the
$i$th and $j$th layer of the relaxed surface, and the lattice
spacing along the (0001) direction of bulk Mg. Table I summarizes
the calculated relaxations for the clean Mg(0001), MgAl1, MgAl2, and
MgAl3 surfaces. We can easily see from Table I that the results
critically depend on the doping length of the Al atom. If the Al
atom is doped at the first (i.e. the topmost) layer, then the
distance between the first and second layers is contracted by 3.00
\%. And if the Al atom is doped at the second layer, then $d_{12}$
and $d_{23}$ respectively decreases by 2.60 \% and 2.12 \%. One can
see that for the layer containing the doped Al atom, the lattice
spacings with its nearest Mg layers are decreased a lot from that of
bulk Mg. This result is confirmed by the relaxation results of the
MgAl3 surface, where $d_{23}$ and $d_{34}$ are decreased by 2.25 \%
and 2.12 \%. By analyzing the atomic charges, we find that the
neighboring Mg atoms always lose electrons to the doped Al atom. So
the charging state of the doping layer and its neighboring Mg layers
are respectively negative and positive, and the interactions between
them become attracting forces. Thus the surface relaxation changes
from expansion of the clean Mg(0001) surface to the contraction
after Al doping.

\begin{table}
\caption{Surface relaxations ($\Delta d_{12}$, $\Delta d_{23}$, and
$\Delta d_{34})$ and work function ($\Phi$) of the Mg(0001), MgAl1,
MgAl2, and MgAl3 surfaces.}
\centering%
\begin{tabular}{c c c c c c}
\hline%
\hline%
 \multirow{2}*{surface} & \multicolumn{3}{c}{Surface relaxation} & \multirow{2}*{$\Phi$ (eV)} \\
\cline{2-4}
 & $\Delta d_{12}$ (\%) & $\Delta d_{23}$ (\%) & $\Delta d_{34}$ (\%) & \\
\hline%
\hline%
Mg(0001) & +1.95 & +0.55 & +0.82 & 3.745 \\
MgAl1    & -3.00 & -0.11 & +0.10 & 3.767 \\
MgAl2    & -2.62 & -2.60 & +0.49 & 3.757 \\
MgAl3    & +0.28 & -2.25 & -2.12 & 3.757 \\
\hline
\end{tabular}\label{TableI}
\end{table}

In addition, we also calculate the work function of the doped
surfaces. The calculated values are given in Table I. We can see
that the doping of Al atoms enlarges the work function by
0.01$\sim$0.02 eV. It means that after doping, surface electrons
become harder to escape into the vacuum, and thus the surface's
reaction activity is lowered.

The band structures of the Mg(0001) and MgAl1 surfaces are
calculated and shown in Fig. 2, together with their electronic
density of states (DOS). By comparing Figs. 2(a) and (b), we can see
that the energy bands around the Fermi energy (E$_f$) change a lot
after Al doping. Through careful wavefunction analysis, we find that
the electronic states between E$_1$ and E$_f$ in Fig. 2(b) mainly
distribute around the top layer. Considering that the electronic
states around E$_f$ are always more important in surface reactions,
the adsorption and dissociation of small molecules on the MgAl1
surface are expected to be different from that on the clean Mg(0001)
surface. As we will see, the dissociation of H$_2$ molecule on the
MgAl1 surface does show different characters.

\begin{figure}[ptb]
\includegraphics[angle=0,width=1.0\textwidth]{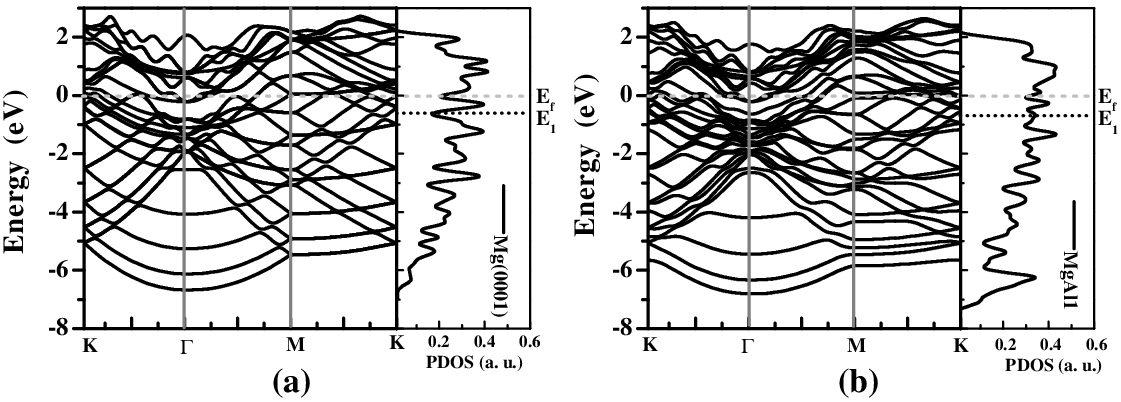}
\caption{(Color online). Band structures and Density of states (DOS)
of the Mg(0001), MgAl1, MgAl2, and MgAl3 surfaces. The Fermi
energies are set to zero.}
\label{fig2}%
\end{figure}

Moreover, we find that for different doping lengths, the result that
the electronic states around E$_f$ distribute around the doping
layer still holds. As shown clearly in Fig. 3, the integral of the
electronic states between E$_1$ and E$_f$ for the MgAl1, MgAl2, and
MgAl3 surfaces respectively have major distributions around the the
first, second and third layers. Since electrons around the first
layer are more easy to take part in surface reactivities, one may
deduce that the influence on surface reactions is weaker for
Al-doping in the second and third layers. And we will prove that it
is true for the dissociation of H$_2$ molecules. By using the Bader
topological method \cite{Bader07} for the MgAl1 surface, we find
that the surface electron density around the first atomic layer also
has a larger distribution around the Al atom, and about 3 electrons
of surface Mg atoms tends to transfer to the Al atom.

\begin{figure}[ptb]
\includegraphics[angle=0,width=0.6\textwidth]{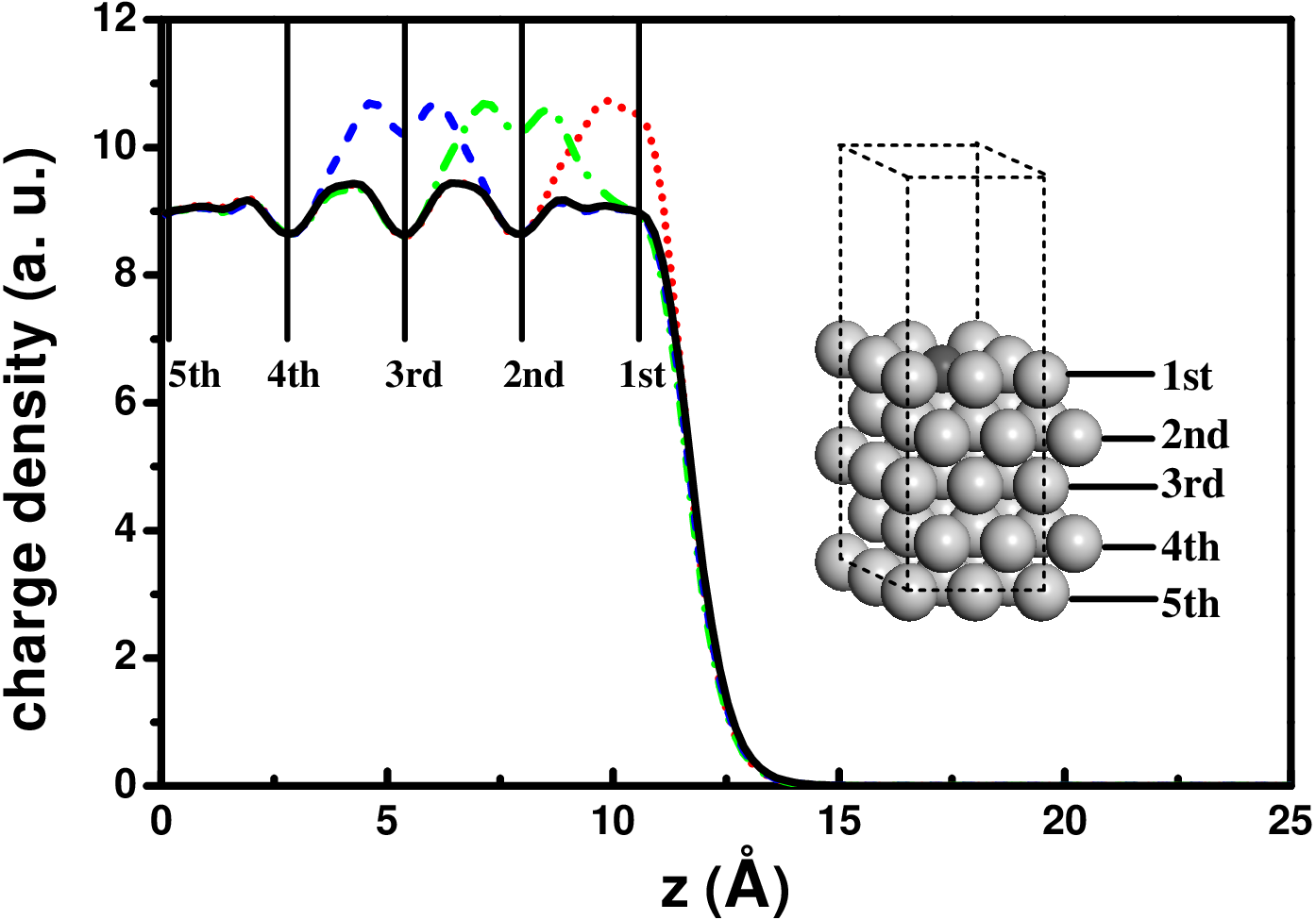}
\caption{(Color online). Band charge distribution around E$_{F}$ of
the clean and Al-incorporation Mg(0001) surface. The energy is set
to [-1.0, 0.1] eV. }
\label{fig3}%
\end{figure}

After systematical studies on the electronic states of the Mg(0001),
MgAl1, MgAl2, and MgAl3 surfaces. We build our model to study the
adsorption of H$_2$ molecules. As shown in Fig. 4(a), around the
doped Al atom there are four different high-symmetry sites on the
surface, respectively, the top, bridge (bri), hcp and fcc hollow
sites. At each site, the adsorbed H$_2$ molecule has three different
high-symmetry orientations, respectively along the $x$ (i.e.,
[$11\bar{2}0$]), $y$ (i.e., [$\bar{1}100$]), and $z$ (i.e.,
[$0001$]) directions. In the following, we will use top-$x,y,z$,
bri-$x,y,z$, hcp-$x,y,z$ and fcc-$x,y,z$ respectively to represent
the 12 high-symmetry channels for the approaching of H$_{2}$.
Through systematic PES calculations, we find that the dissociation
energy barrier of H$_{2}$ along other low-symmetry channels is
always larger than along these high-symmetry channels, similar to
that observed during the dissociation of oxygen molecules on the
Be(0001) surface \cite{Zhang09}.

\begin{figure}[ptb]
\includegraphics[angle=0,width=0.6\textwidth]{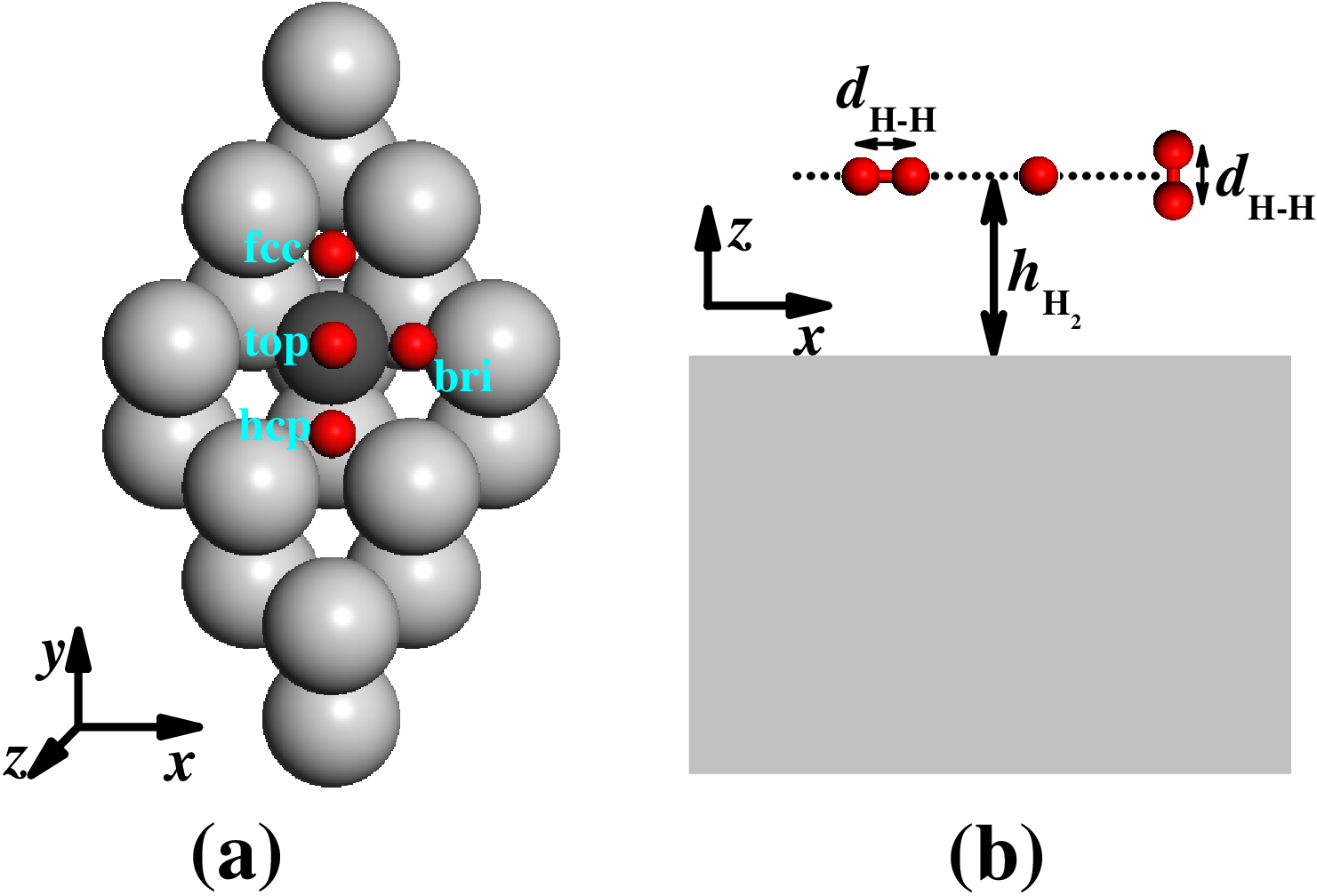}
\caption{(Color online). (a) The $p$($2\times2$) surface cell of
Mg(0001) (clean or Al-doped) and four on-surface adsorption sites.
(b) The sketch map showing that the molecule (with vertical or
parallel orientation) is initially away from the surface with a
hight $h_{\mathrm{H_{2}}}$. Mg and Al atoms are shown in grey and
dark grey balls. The H$_2$ molecule is depicted by red balls.}
\label{fig4}%
\end{figure}

By calculating the two-dimensional (2D) potential energy surface
(PES) cuts, we find that the minimum energy path for dissociation of
H$_2$ on the Mg(0001) surface is along the bri-$y$ channel, in
accordance with previous reports \cite{Vegge04,Johansson06,Wu08}.
And the corresponding energy barrier ($\Delta E$) is 0.85 eV. After
doping of an Al atom, we find that the dissociation path with the
lowest energy barrier is still along the bri-$y$ channel on the
MgAl1, MgAl2, and MgAl3 surfaces. The calculated energy barriers,
however, are all found to be larger than that on a clean Mg(0001)
surface. As a typical prototype, the PES cut for a H$_2$ molecule
along the bri-$y$ channel on the MgAl1 surface is shown in Fig. 5.
And the calculated energy barriers for the high-symmetry adsorption
channels on the four surfaces are summarized in Table II, together
with the bond length and height of the H$_2$ at the corresponding
transition states. We can see from Table II that as the doping
length of the Al atom is enlarged, the dissociation energy barrier
approaches to the value of clean Mg(0001) surface. So the doped Al
atom needs to be closer to the surface to make dissociation of H$_2$
harder.

Moreover, we can also see from Table II that for the bri-$y$
channel, the molecular height for the transition state is smaller on
the MgAl1 (1.04 \AA) than on the clean Mg(0001) surface (1.15 \AA).
As we have already seen, the work function of the Mg(0001) surface
is enlarged after Al-doping, and surface electrons become harder to
escape into the vacuum. Since that a H$_2$ molecule needs two more
electrons to dissociate into two hydrogen atoms, the fact that a
H$_2$ molecule needs to be closer to the MgAl1 surface to dissociate
is a chemical reflection of the change of the surface electronic
states.

\begin{figure}[ptb]
\includegraphics[angle=0,width=0.6\textwidth]{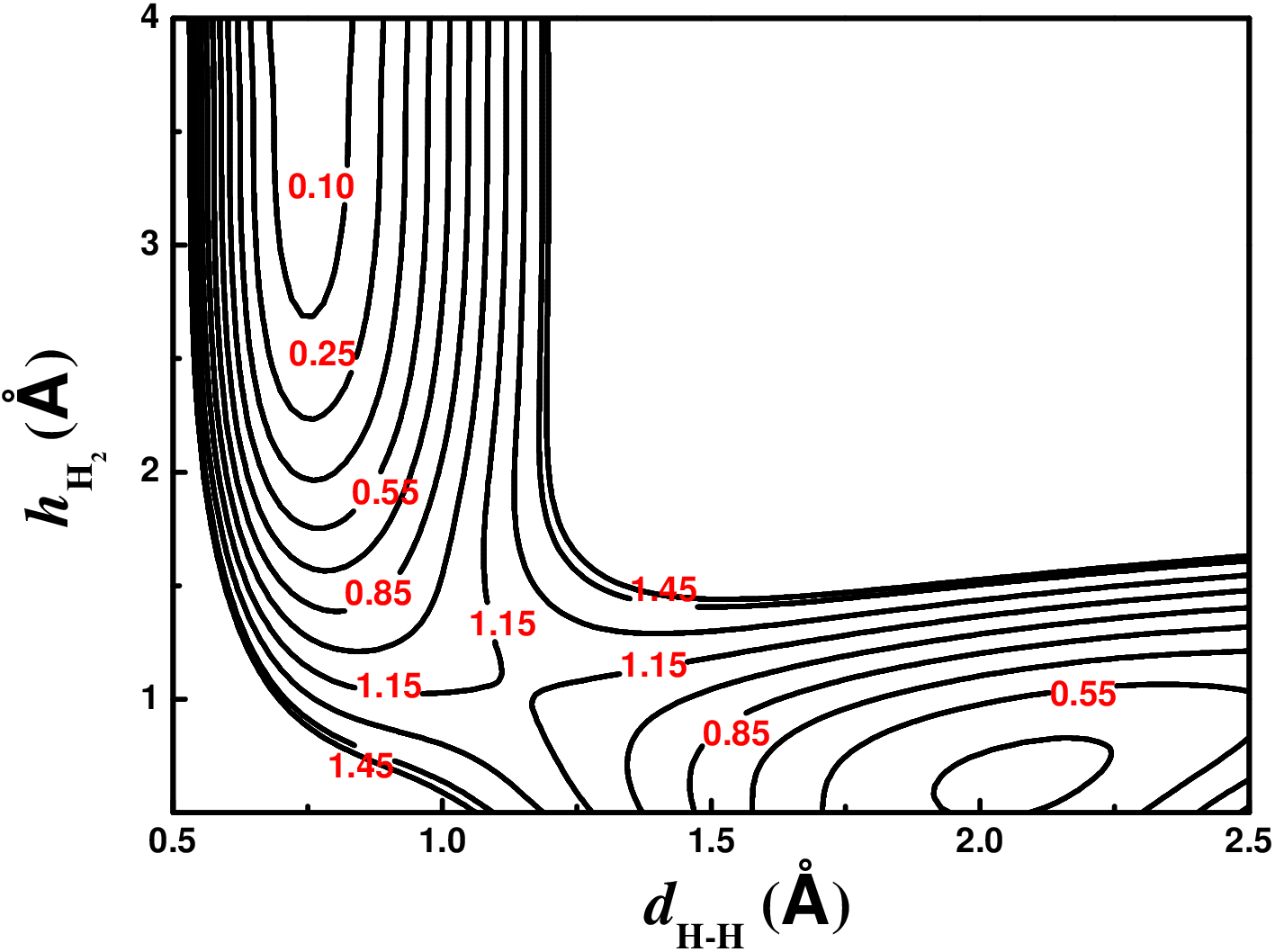}
\caption{(Color online). Contour plots of the two dimensional cuts
of the potential energy surfaces (PESs) for dissociation of H$_{2}$
on the MgAl1 surface, as a function of the bond lengths
($d_{\mathrm{H-H}}$) and the heights ($h_{\mathrm{H_{2}}}$) along
the bri-$y$ adsorption channel.}
\label{fig5}%
\end{figure}

\begin{table}[ptb]
\caption{Energy barrier, bond length and height of the H$_2$
molecule at the transition states along the 12 high-symmetry
adsorption channels on the Mg(0001), MgAl1, MgAl2 and MgAl3
surfaces. Energy barriers are in units of eV, while bond lengths and
heights are in units of \AA.}
\centering%
\begin{tabular}[c]{ccccccccccccc}
\hline%
\hline%
Adsorption & \multicolumn{3}{c}{clean Mg(0001)} & \multicolumn{3}{c}{MgAl1} & \multicolumn{3}{c}{MgAl2} & \multicolumn{3}{c}{MgAl3} \\
\cline{2-13}%
channel    & barrier & $d$(TS) & $h$(TS) & barrier & $d$(TS) & $h$(TS) & barrier & $d$(TS) & $h$(TS) & barrier & $d$(TS) & $h$(TS) \\
\hline%
\hline%
bri-x & 1.60 & 1.70 & 1.64 & 1.42 & 1.54 & 1.44 & 1.64 & 1.65 & 1.52 & 1.66 & 1.62 & 1.60 \\
bri-y & 0.85 & 1.12 & 1.15 & 1.15 & 1.14 & 1.04 & 0.91 & 1.07 & 1.11 & 0.90 & 1.08 & 1.12 \\
top-x & 1.62 & 1.58 & 1.64 & 1.35 & 1.29 & 1.38 & 1.63 & 1.57 & 1.55 & 1.61 & 1.52 & 1.62 \\
top-y & 1.63 & 1.58 & 1.64 & 1.35 & 1.29 & 1.36 & 1.65 & 1.57 & 1.54 & 1.65 & 1.54 & 1.62 \\
hcp-x & 1.25 & 1.34 & 1.30 & 1.24 & 1.23 & 1.19 & 1.24 & 1.31 & 1.20 & 1.21 & 1.36 & 1.28 \\
hcp-y & 1.26 & 1.36 & 1.31 & 1.48 & 1.54 & 1.28 & 1.25 & 1.32 & 1.22 & 1.26 & 1.36 & 1.28 \\
fcc-x & 1.22 & 2.33 & 1.30 & 1.22 & 2.08 & 1.16 & 1.34 & 2.27 & 1.30 & 1.20 & 2.28 & 1.28 \\
fcc-y & 1.23 & 2.33 & 1.30 & 1.51 & 2.48 & 1.32 & 1.31 & 2.39 & 1.27 & 1.23 & 2.40 & 1.28 \\
\hline%
\end{tabular}\label{TableII}
\end{table}

For the dissociation of H$_2$ molecules on most metal surfaces, it
has been revealed that two different interactions exist, which are
respectively the orthogonalizations between molecular orbitals of
H$_2$ and electronic states of metal, and the charge transfer from
metal atoms to hydrogen atoms \cite{Harris1985,Hammer1993,Li09CPB}.
As shown in Fig. 2(b), we find that the Al-doping did not change the
metallic properties of the Mg(0001) surface. So such two different
interactions still exist for the dissociation of H$_2$ on the MgAl1
surface.

\begin{figure}[ptb]
\includegraphics[width=0.8\textwidth]{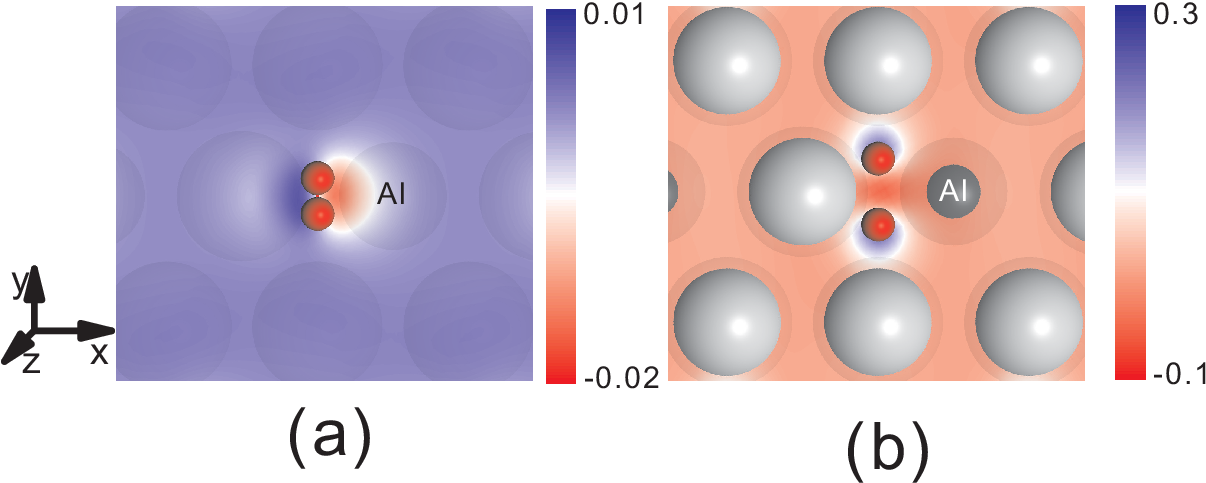}
\caption{(Color online). The difference electron density for the
H$_2$/MgAl1 system with the height of H$_2$ to be 2.00 (a) and 0.92
\AA~ (b). Mg, Al, and H atoms are shown in grey, dark grey, and red
balls respectively. All the values are in unit of electrons per
\AA$^3$.}
\label{fig6}%
\end{figure}

Figures 6(a) and (b) respectively show the difference electron
density for the adsorption system of H$_2$/MgAl1 before ($h_{\rm
H_2}$=2.00 \AA) and after ($h_{\rm H_2}$=0.92 \AA) the transition
state along the bri-$y$ channel, namely,
\begin{equation}\label{Drho}
\Delta\rho =\rho({\rm H_2+ MgAl1}) - \rho({\rm H_2}) - \rho({\rm
MgAl1})],
\end{equation}
where $\rho({\rm H_2+ MgAl1})$, $\rho({\rm H_2})$ and $\rho({\rm
MgAl1})$ are respectively the electron density of the adsorption
system, the H$_2$ molecule and the MgAl1 surface. To calculate
$\Delta\rho$, the atomic positions in the last two terms in Eq. 2
have been kept at those of the first term. Through careful
wavefunction analysis, we find that at the beginning of the
adsorption process, the molecular orbitals of H$_2$ orthogonalize
with electronic states of the MgAl1 surface and are thus broadened.
This orthogonalization is very similar to that has been observed
during the interaction of H$_2$ with the Mg(0001) \cite{Li09CPB} and
Al(111) \cite{Hammer1993} surfaces. As shown in Fig. 6(a), there is
an electron-depletion region on top of the Al atom, which is formed
because electrons around the Al atom are repelled by the H$_2$
bonding electrons due to the orthogonalization. In comparison, on
the clean Mg(0001) surface, the electrons of Mg are repelled from
the surface at the beginning of the adsorption of a H$_2$ molecule.
We can also see from Fig. 6(b) that the H$_2$ molecule begins to get
some electrons from the Mg and Al atoms, when it gets more closer to
the MgAl1 surface.

As we have already seen, extra electrons on the MgAl1 surface trends
to distribute around the Al atom. Thus during the initial stage of
the adsorption of a H$_2$ molecule, the molecular orbitals of H$_2$
orthogonalize only with electronic states around the Al atom, while
on the clean Mg surface, the molecular orbitals orthogonalize with
electronic states around the two nearest Mg atoms. Besides, the
ability to lose electrons is stronger for a Mg atom than an Al atom.
Therefore, the H$_2$ molecule is more difficult to get electrons
from the MgAl1 surface. And so the dissociation energy barrier is
larger on the MgAl1 surface.

\section{Summary}

In summary, the electronic structures of the Al-doped Mg(0001)
surfaces and the dissociation behaviors of H$_2$ molecules on the
doped surfaces are systematically studied, using the
first-principles calculations. Our calculational results show that
after Al-doping, the surface electronic structure is largely
changed. Due to the electrons redistribution, the surface relaxation
changes from expansion to contraction around the doping layer. For
different doping lengths, the work function of the Mg(0001) surface
are all enlarged. Through careful electronic structure analysis, we
further find that the electronic states around the Fermi energy
mainly distribute around the doping layer for Al-doped Mg(0001)
surface. By calculating the potential energy surface of a H$_2$
molecule on the Al-doped Mg(0001) surfaces, we find that the doping
of an Al atom does not change the dissociation path with the lowest
energy barrier, but enlarges the energy barrier. And for the first
layer doping, the energy barrier is changed by 0.30 eV. From
wavefunction and charge-density analysis, we reveal the two
interaction stages during the dissociation of the H$_2$ molecule,
i.e. the first orthogonalization and later charge-transfer stages.
We find that because more surface electrons distribute around the Al
atom, the orthogonalization mainly happens between the H$_2$
molecule and the surface Al atom. In addition, since Al atom has a
weaker ability to lose electrons, the dissociation energy barrier
for H$_2$ is enlarged on the MgAl1 surface than on a clean Mg(0001)
surface. Our studies provide a detailed mechanism for explaining the
improvement on the resistance to corrosion for the Mg(0001) surface
by doping of Al atoms.

\begin{acknowledgments}
P. Z. was supported by the NSFC under Grants No. 10604010 and No.
60776063. Y. W. was supported by the NSFC under Grants No. 50471070
and No. 50644041.
\end{acknowledgments}

\end{document}